\documentclass{jpsj2}

\def\runauthor{Ke-Ichi Fushimi et al.}

\title{%
WIMPs search by means of the highly segmented scintillator
}

\author{%
K.Fushimi\thanks{E-mail address: fushimi@ias.tokushima-u.ac.jp}$^{\rm{a}}$
H.Kawasuso$^{\rm{a}}$ M.Toi$^{\rm{a}}$,E.Aihara$^{\rm{a}}$,R.Hayami$^{\rm{a}}$,
S.Nakayama$^{\rm{a}}$,N.Koori$^{\rm{a}}$,R.Hazama$^{\rm{b}}$,
K.Ichihara$^{\rm{b}}$,Y.Shichijo$^{\rm{b}}$,
S.Umehara$^{\rm{b}}$ and S.Yoshida$^{\rm{b}}$
}

\inst{%
(a) Faculty of Integrated Arts and Sciences, The University of 
Tokushima, 1-1 Minami Josanjimacho Tokushima city, Tokushima 770-8502, JAPAN \\
(b) Department of Physics, Osaka University, 1-1 Machikaneyamacho Toyonaka
city, Osaka 560-0047, JAPAN
}

\recdate{\today}

\abst{%
The highly sensitive method to search for WIMPs dark matter particles is 
proposed.
An array of thin NaI(Tl) plate has the great selectivity
for distinguishing the WIMPs events and background ones.
The principle of signal selection for WIMPs is described.
The high sensitivity for SD (spin-dependent) type WIMPs is expected 
by applying multi-layer system of NaI(Tl) detector.
}

\kword{%
Thin scintillator, WIMPs search, Dark matter
}

\begin{document}
\maketitle

\section{Introduction}
A method of highly sensitive measurement is strongly needed to study
nuclear and particle rare processes.
The need for highly sensitive radiation detector has been increased in the 
fields of particle astrophysics and nuclear physics.
Particularly, the search for dark matter candidates and the search for 
neutrino-less double beta decay need to select the signal events from the 
huge number of background ones.
In the case of WIMPs search, the groups such as ELEGANT \cite{fushimi},
DAMA \cite{DAMA} applied the huge volume of NaI(Tl) scintillator.
The new project with the huge volume scintillator are proposed 
with hundreds kg of scintillators by XMASS\cite{XMASS} group,
ZEPLIN group \cite{Zeplin} and so on.

The many groups are trying to enhance the sensitivity by developing a 
large volume and low background detectors.
However, to reduce the background events becomes more difficult 
for the larger detectors because of its poor position resolution.
We propose highly selective detector system to extract the 
signals of WIMPs.
The segmentation of a detector results good position resolution, 
which enhances the sensitivity for WIMPs.

The three types of interaction between WIMPs and nucleus has been 
proposed by many theorists.
One is spin-independent (SI), the second 
is spin-dependent (SD), and the third one is spin-dependent inelastic 
scattering (EX).
The SI interaction cross section depends on the square of nuclear mass 
number $A^{2}$.\@
On the other hand, the SD one depends on the nuclear spin-matrix element
$\lambda^{2}J(J+1)$.\@
The EX interaction cross section is strongly related to the 
spin-dependent (SD) interaction cross section.
Consequently, the event rate of EX and SD are related each other,
moreover, the relationship is enhanced by measuring the annual modulation.
The highly selective detector which measures both EX and elastic scattering
contributes much for the WIMPs search.

It consists of an array of thin NaI(Tl) scintillator plate.
The event selection power is largely enhanced by applying the 
compact modular detector system.
The size of each modules is designed in order to enhance the discrimination 
between the signal and the background.

\section{Signal Selection by Space and Time Correlation (SSSTC)}
The WIMPs signal is expected to make the localized event both in space and 
in time.
On the other hand, the background events make the diffused and correlated ones.
Consequently, the spatial correlation and timing correlation is 
quite important information to extract the signal event from the large 
number of background events.
\begin{figure}[h]
\includegraphics[width=14cm]{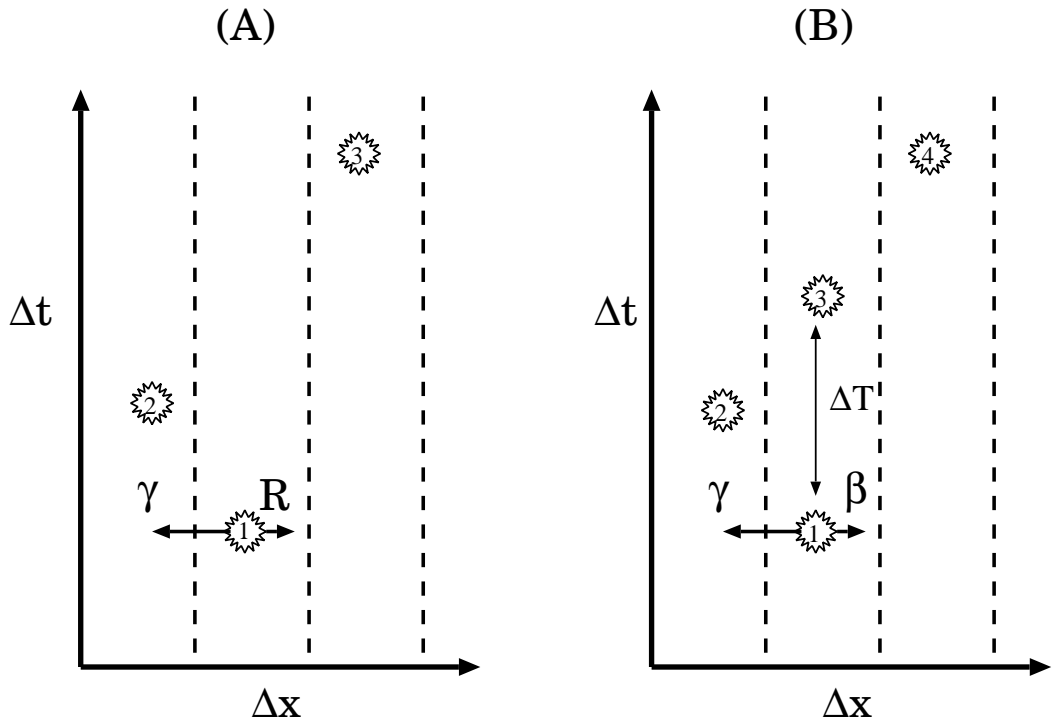}
\caption{
Schematic drawing of the principle of SSSTC. A detector is spatially segmented 
by the vertical dashed lines. }
\label{fg:ssstc}
\end{figure}
Fig.\ref{fg:ssstc}(A) shows the schematic spatial-timing correlation of 
the event in the case of WIMPs-nuclear inelastic excitation.
The number 1,2,3 and 4 stand for the order of events occurred.
The event indicated ``1'' is the signal event for WIMPs-nucleus interaction,
and the ones ``2'' and ``3'' are background events.

In the case of elastic scattering, the events cannot be the spatially diffused 
event, and must be the timing isolated event because the range of recoil 
nucleus (only ``R'' in Fig.\ref{fg:ssstc}) is much shorter than the thickness 
of each module.

In the case of inelastic excitation of $^{127}$I nucleus, 
the 57.6keV $\gamma$ ray is simultaneously emitted.
Such the signal events for WIMPs-nucleus interaction produce the characteristic
events.
The recoil nucleus ``R'' produces a continuous energy spectrum in the low 
energy region and the low energy $\gamma$ ray ``$\gamma$'' in Fig.\ref{fg:ssstc}
is observed in the another modules.

Fig.\ref{fg:ssstc}(B) shows that the RI background event which makes a similar 
event.
The $\beta-\gamma$ successive transition makes similar event to the 
signal ones.
The most serious origin of such a fake event is due to $^{210}$Pb.
The radioactivity of $^{210}$Pb in NaI(Tl) was about 10mBq/kg, which is 
one thousand times larger than that of $^{226}$Ra\cite{ELEV}.
The background events due to $^{210}$Pb is tagged by the following $\beta$
decay event by $^{210}$Bi and eliminated.
The accidental background events suffers the event selection in the 
case of a large volume detector.
Many groups has tried to achieve good position resolution by means of 
pulse shape discrimination or timing discrimination or pulse height dependence.
However, it is quite clear that the segmentation of the detector 
is achieved most effectively by means of small modules of detector.
In the following sections, the effectiveness of event selection 
of WIMPs-nucleus inelastic excitation (section \ref{sec:signal})
and the reduction power of $^{210}$Pb background (section \ref{sec:210pb})
by means of segmentation.

\section{Event selection by means of segmented NaI(Tl)}
\label{sec:signal}
In this section, the selectivity of EX events by highly segmented scintillator
is discussed.
Since the attenuation coefficient of 57.6keV $\gamma$ ray is 
as large as 6.83cm$^{2}$/g, each detector module must be sufficiently small.
We determined the design of the NaI(Tl) scintillator so as to 
detect the $\gamma$ ray by the next module with enlarging the volume 
of the detector.
The NaI(Tl) crystal with 5cm$\times$5cm in area and less than 
2mm in thickness was considered by performing Monte Carlo simulation.

\begin{figure}[h]
\includegraphics[width=10cm]{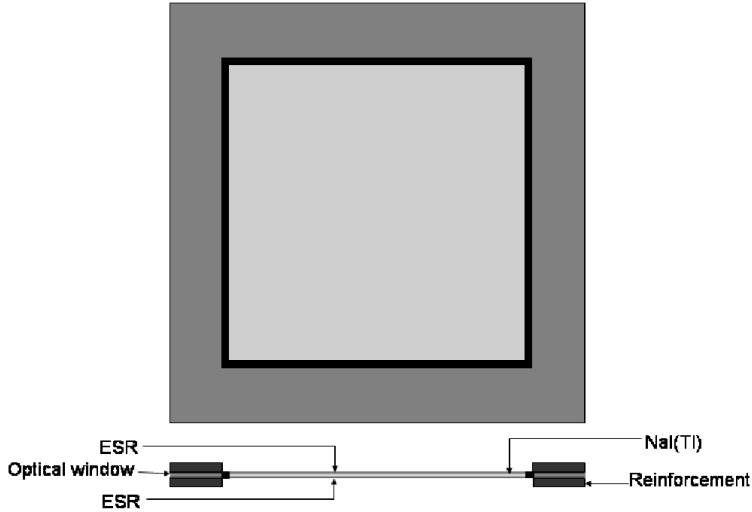}
\caption{
The schematic drawing of a NaI(Tl) module.
}
\label{fg:nai}
\end{figure}

The conditions which has been considered in the simulation are listed below.
\begin{enumerate}
\item The thickness of NaI(Tl) crystal was varied 0.1mm to 1.5mm.
\item The wider area of the NaI(Tl) crystal was covered with ESR$^{TM}$
(Enhanced Specular Reflector)reflector sheets.
\item Four light guides made of acrylic plate were placed on the thinner 
sides of NaI(Tl) crystal.
\item A reinforcement was glued on ESR$^{TM}$ sheet.
\item Ten modules were piled up.
\item The inelastic excitation occurred in the 5th module.
\item The simulation was performed with 1 million events. for EX and 
$^{210}$Pb, 3 million events for $^{210}$Pb.
\item The energy resolution of the thin NaI(Tl) was 
assumed as large as 19\% in FWHM (Full Width Half Maximum) at 60keV.
\end{enumerate}
The drawing of one module is shown in Fig.\ref{fg:nai}.
Optical cement is glued between NaI(Tl) and optical window.

\begin{figure}[h]
\includegraphics[width=10cm]{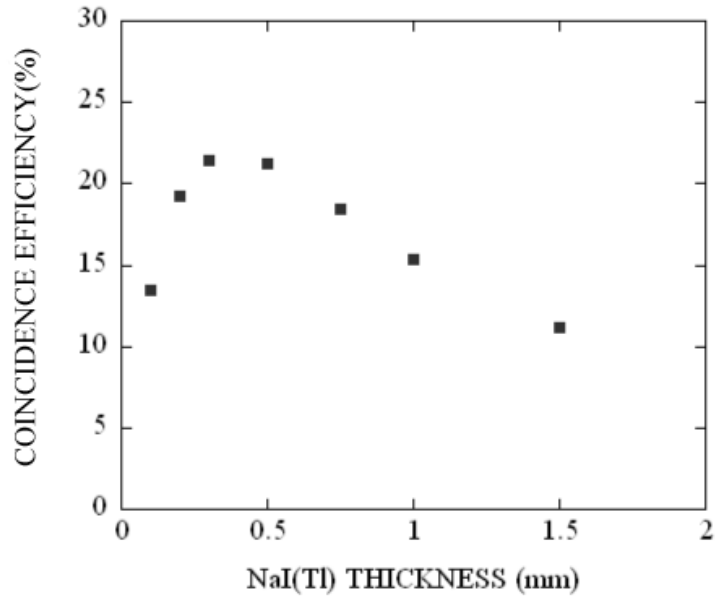}
\caption{
The thickness dependence of coincidence efficiency.
}
\label{fg:eff-thick}
\end{figure}

The analysis was performed in order to extract the EX events.
The number of the events in the energy range between 47keV and 67keV
($\simeq$57.6keV$\pm$10keV)
which were detected by 4th or 6th module was divided by the number of total 
simulated events (1 million events) to estimate the 'coincidence efficiency'.
Fig.\ref{fg:eff-thick} shows the thickness dependence of coincidence 
efficiency.

The coincidence efficiency has a maximum between 0.3mm and 0.5mm.
In the case of the NaI(Tl) is thinner than 0.3mm, the $\gamma$ ray 
goes through the next module, which makes a spatially diffused event.
On the other hand, the coincidence efficiency does not decrease rapidly 
when the NaI(Tl) becomes thicker.
However, since the thicker module makes larger event rates 
of background events,
the background events suffers the sensitivity.
The background reduction power will be discussed in the next section.

Recently, the suitable thin NaI(Tl) crystal with the wide area in 
collaboration with Horiba Ltd.
The thin NaI(Tl) detector whose dimension was 5cm$\times$5cm$\times$0.05cm
was successfully developed\cite{thinnai0}.
The thin NaI(Tl) plate with 0.05cm in thickness is suitable 
for SSSTC measurement of WIMPs search.
The performance of the thin NaI(Tl) was tested by low energy photons.
The good energy resolution of 19\% at 60keV and low energy threshold 
of about 2keV was obtained by thin NaI(Tl) detector\cite{thinnai0}.

\section{Background reduction by means of segmented NaI(Tl)}
\label{sec:210pb}
The background events for WIMPs search are mainly due to the internal 
RI's (radioactive isotopes) which are contained in the sensitive volume 
of the detector.
The external RI's which are contained in the surrounding materials also 
make serious background.
Since the background events due to external RI's are mainly due to Compton 
scattering of high energy $\gamma$ rays, 
the background events are effectively reduced by the highly segmented 
detector system.
Moreover, the internal RI's are much more serious than external
ones because of the large geometrical efficiency.
Consequently, the background events due to only the internal RI's were 
considered in the present estimation.

The most serious origin of background is $^{210}$Pb.
$^{210}$Pb emits low energy $\beta$ ray ($E_{max}=16.5$keV) and low energy 
$\gamma$ ray ($E_{\gamma}=46.5$keV).\@
The simultaneous emission of 
the low energy $\beta$ ray and the $\gamma$ ray makes a similar event
to a signal event of EX. 
The branching ratio that $^{210}$Pb decays to the first excited state is 84\%.
The transition from the first excited state to the ground state 
is mainly by internal conversion, consequently, the probability of 
$\gamma$ ray emission is 4.25\%.

The normal NaI(Tl) crystal contains a large amount of $^{210}$Pb.
For example, the concentration of $^{210}$Pb was 8.4mBq/kg in the NaI(Tl) 
of ELEGANT V,
while the one of $^{214}$Pb which is the parents of $^{210}$Pb was 
as small as a few tens of $\mu$Bq/kg.
The radiation equilibrium in U-chain in NaI(Tl) is often broken 
\cite{umehara,ichihara}.
The progeny after $^{222}$Rn in U-chain were concentrated from the 
air, water and raw materials.

It is quite difficult to reduce the concentration of $^{210}$Pb, however, 
its contribution to background is efficiently reduced by SSSTC analysis.
In the segmented detector module, the progeny of $^{210}$Pb decays after 
its decay and detected in the same module.
Suppose the 1kg detector is divided into $n$ modules which contains 
$b$Bq/kg of $^{210}$Pb.
The decay rate in one module is simply expressed as 
$b/n$sec$^{-1}$module$^{-1}$.
The sequential events which occur during the interval $\Delta T$ 
are rejected as the decay chain of 
$^{210}$Pb$\rightarrow^{210}$Bi$\rightarrow^{210}$Po.
In this case, the accidental decay due to another $^{210}$Pb 
must not occur, thus the mean time interval due to $^{210}$Pb  
must be longer than the time interval. 
Thus, the condition
\begin{equation}
\Delta T < \frac{n}{b} 
\label{eq:con}
\end{equation}
is needed.
The reduction power for BG events is enhanced with setting a 
longer time interval $\Delta T$.
The time interval $\Delta T=N\times T_{1/2}$, with $N=2\sim3$
is adequate for SSSTC analysis.

The present WIMPs search project aims to develop the highly pure 
NaI(Tl) crystal whose concentration of $^{210}$Pb is as small as 
0.1mBq/kg.
The target concentration of $^{210}$Pb is adequately realistic 
purity.
In this case, segmentation $n$ is calculated as 
\begin{eqnarray}
n & > & bNT_{1/2} \\ \nonumber
  & = & 1\times10^{-4}\cdot(2\sim3)\cdot4.33\times 10^{5} \\ \nonumber
  & = & 87\sim130{\rm \ segmentation/kg}.
\end{eqnarray}
The proposed NaI(Tl) module whose dimension of 5cm$\times$5cm$\times$0.05cm
corresponds to 217 segmentation.
The high sensitive measurement is achieved by the highly segmented 
detector system with the feasible purity of background radioactivity.

\section{Estimation of the sensitivity for inelastic excitation}
The expected sensitivity for inelastic scattering of WIMPs-$^{127}$I
will be discussed in this section.
The expected background events were estimated by means of Monte Carlo 
simulation.
The serious background origins, $^{214}$Pb, $^{214}$Bi, $^{210}$Pb and 
$^{210}$Bi in were considered.
The contribution of $^{40}$K is expected to be small because of extremely 
small radioactivity (less than 0.2ppm in $^{nat}$K) in the present NaI(Tl).
The simulated concentration of $^{214}$Pb and $^{214}$Bi was 0.01mBq/kg 
which was precisely measured by delayed coincidence method.

While the simulated concentration of $^{210}$Pb was 0.1mBq/kg, 
which was 1/100 smaller than the present concentration.
The radioactive equilibrium in U-chain often breaks because 
of the concentration of $^{222}$Rn in the air. 
Our previous work found that the concentration of $^{210}$Pb is 
almost three orders of magnitude larger than the one of $^{214}$Pb and
$^{214}$Bi.
The contamination of $^{210}$Pb is reduced by controlling the air
during the creation processes of NaI(Tl) crystal.
The Monte Carlo simulation was performed under the condition that 
0.1mBq/kg in NaI(Tl) crystal.

The SSSTC analysis was performed for the simulated background data.
A event in the energy region 57.6keV$\pm$10keV was selected for the 
inelastic excitation of $^{127}$I.\@
After the analysis, the accidental coincidence events were reduced 
by setting the timing correlation.
The successive events during the specific time were concluded to be 
the decays belonging to the decay chain.
In the case of the chain between $^{214}$Pb and $^{214}$Bi, 
the event which occurs 1 second and 3600 seconds after the preceding event
is removed as the background one.
The fraction of removed events is estimated as 
\[
\int^{3600}_{1}\left(\frac{1}{2}\right)^{\frac{t}{T_{1/2}}}dt=0.997.
\]
In the case of the decay chain from $^{210}$Pb to $^{210}$Po, the 
82.3\% of the background events are removed with setting the 
time difference being 12.5days (2.5$T_{1/2}$).

\begin{figure}[ht]
\includegraphics[width=10cm]{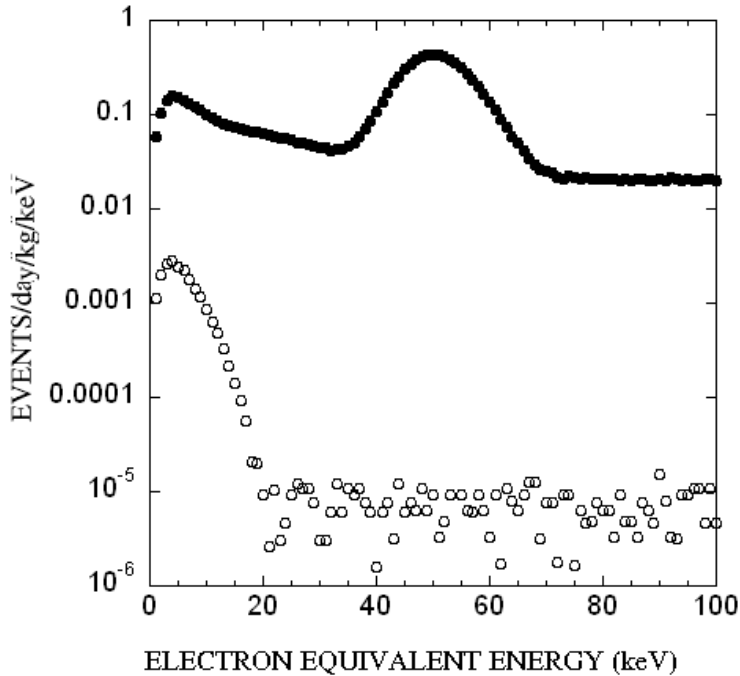}
\caption{
The expected background energy spectrum.
The closed circles and open circles mean the singles event rate 
and the event rate after performing the SSSTC analysis.
The analysis process is described in text.
}
\label{fg:est-spe}
\end{figure}

The estimated background energy spectrum is shown in Fig.\ref{fg:est-spe}.
The energy resolution which was measured by actual thin NaI(Tl) was 
considered in this simulation.
The events between 2keV and 10keV was summed to estimate the sensitivity for 
WIMPs-nucleus inelastic excitation.
The energy spectrum after performing the SSSTC analysis has the big 
contribution by $^{210}$Pb. 
A big bump below 20keV of the open circle is due to the beta ray 
from $^{210}$Pb ($E_{max}=16.5$keV).

The expected event rate due to the radioactive contamination which is contained
in the NaI(Tl) crystal was calculated as 1.7$\times 10^{-2}$/kg/day.
The assumed energy threshold was 2keV in electron equivalent and the 
event rate between 2keV and 10keV was integrated.
The energy threshold, 2keV, was already achieved by thin NaI(Tl) scintillator
\cite{thinnai0}.
The statistical accuracy is improved as the increase of the modules.
The expected background rate and the upper limit (90\% C.L.) 
of the event rate are listed in table \ref{tb:limit}.
\begin{table}[ht]
\caption{
The expected event rate for multi layer detector system of NaI(Tl) plate.
The upper limit on the event rate at 90\% C.L. are shown.
}
\label{tb:limit}
\begin{tabular}{rrrr} \\ \hline 
Number of modules & Total mass(kg) & BG rate(kg$^{-1}$y$^{-1}$) & 
Upper limit ($R_{lim}$kg$^{-1}$day$^{-1}$) \\ \hline
16 & 0.0734 & $0.45\pm0.67$& $5.2\times 10^{-2}$ \\
256& 1.17   & $7.2\pm 2.7$ & $8.1\times 10^{-3}$ \\
1024&4.70   & $28.9\pm 5.4$& $4.0\times 10^{-3}$ \\
2176&9.98   & $61.3\pm 7.8$& $2.8\times 10^{-3}$ \\ \hline
\end{tabular}
\end{table}

The expected sensitivity for the cross section of inelastic excitation 
is calculated easily by the formula,
\begin{equation}
\sigma_{EX,lim}=\frac{m_{\chi}}
{N_{T}\rho_{0}\left< v\right>f\left|F(q)\right|^{2}\epsilon}\times R_{lim}.
\end{equation}
Where $m_{\chi}$ and  $N_{T}$ are the mass of WIMPs and the number density of 
target nucleus ($^{127}$I), $\rho_{0}=0.3$GeV/cm$^{3}$ is the local 
halo density, $\left< v\right>=230$km/sec is the mean velocity of WIMPs.
The phase space factor $f=p_{f}/p_{i}$ was calculated numerically.
The form factor $F(q)$ was used for the spin-dependent form factor
\cite{Ressell}.
The coincidence efficiency $\epsilon=0.21$ is discussed in the previous 
section.

The other background from out of the detector should be considered.
The background is mainly due to high energy gamma rays from photomultiplier 
tubes (PMT).
The high energy gamma rays are reduced because they interact in many 
modules through multiple Compton scattering.
Since the multiple Compton scattering events are effectively reduced,
the BG's due to external origins of the detector is ignored in the 
present report.

The cross sections of EX and SD are closely related by the nuclear spin 
matrix element.
The case of elastic scattering, the nuclear spin-matrix element, 
$\lambda^{2}J(J+1)$ has the large model dependence \cite{Ressell}.
The spin-matrix element using the calculation by means of 
single particle shell model was applied to our estimation\cite{Ressell}.
However, the spin-matrix element for $^{127}$I cancels out when one 
renormalizes the sensitivity to WIMPs-proton cross section.
The spin-matrix element for proton has no model dependence and the 
value is $\lambda^{2}J(J+1)=0.75$.

The case of inelastic scattering the nuclear spin-matrix element,
$\frac{2J^{\ast}+1}{2J+1} \left| \frac{M_{M1}}{\mu_{p}}\right|$
is experimentally deduced from the nuclear transition probability,
where $J$ and $J^{\ast}$  are the total angular momentum of 
the ground state and excited state, and $\mu_{p}$ is 
the magnetic dipole moment of proton.
Consequently the precise exclusion plot with small model dependence 
is obtained.
The sensitivity of the cross section $\sigma_{lim,p-\chi}$ is 
calculated simply,
\begin{equation}
\sigma_{lim,p-\chi}=\left[\lambda^{2}J(J+1)\right]_{p}
\frac{m_{p}^{2}}{m_{N}^{2}}\left(\frac{m_{p}+m_{\chi}}
{m_{N}+m_{\chi}}\right)^{2}
\left|\frac{\mu_{p}}{M_{M1}}\right|^{2}
\frac{2J^{\ast}+1}{2J+1}\sigma_{lim,EX}.
\end{equation}
In the present estimation, the parameters in the formulae are 
listed in table \ref{tb:param}.
\begin{table}[ht]
\caption{The parameters which are used for the present estimation.}
\label{tb:param}
\begin{tabular}{lrl}\hline
Parameter   & Value  & Ref. \\ \hline
$m_{p}$    & 938GeV & \cite{TOI} \\
$m_{N}$    & 118.18GeV & \cite{TOI} \\
$J$        & 5/2       & \cite{TOI} \\
$J^{\ast}$  & 7/2       &  \cite{TOI} \\
$\left|M_{M1}\right|^{2}$& 0.10 & \cite{Ellis} \\
$\mu_{p}$  & 2.79 & \cite{TOI}\\ \hline
\end{tabular}
\end{table}

\begin{figure}[ht]
\includegraphics[width=10cm]{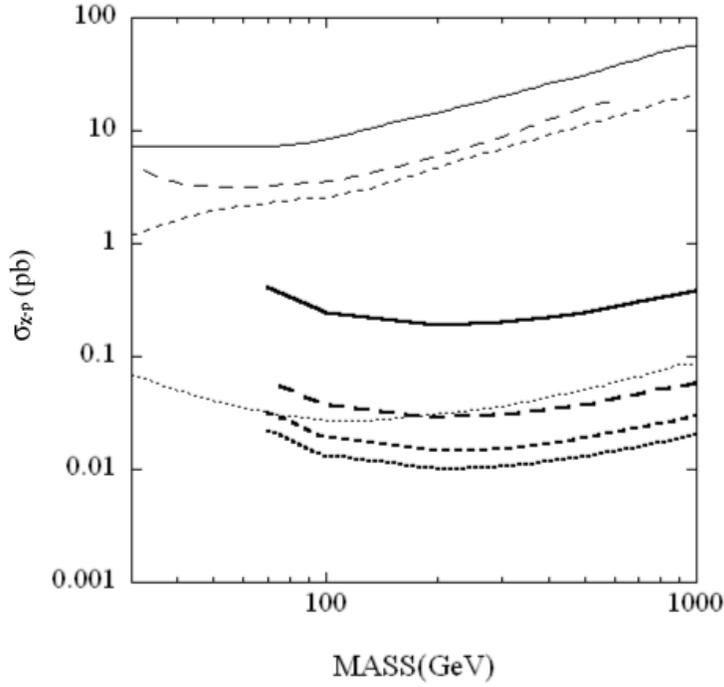}
\caption{
Thick lines are the expected sensitivity for SD type WIMPs.
The solid, long-dashed, sort-dashed and dotted lines are the 
expected sensitivity by 16, 256, 1024 and 2176 modules of NaI(Tl) array.
Thin lines are results of CRESST (solid line), DAMA LXe (long-dashed line),
ELEGANT V at OTO (short-dashed line) and the expected sensitivity by 
NAIAD (dotted line).
}
\label{fg:sens}
\end{figure}

The estimated sensitivity for SD type WIMPs is shown in Fig.\ref{fg:sens}.
The high sensitivity will be expected by the small amount of 
NaI(Tl) crystal.

\section{Concluding remarks and future prospect}
The WIMPs search by means of highly segmented NaI(Tl) scintillator
was proposed in the present work.
It is important to investigate both the elastic scattering and 
the inelastic scattering of WIMPs-nucleus interaction.
The ratio between elastic scattering and inelastic scattering 
provides the quite important information for WIMPs nature.
We showed the thin and wide area NaI(Tl) plate was suitable for 
study the WIMPs interaction.

Since NaI(Tl) scintillators have been well developed, 
the production of highly pure and large volume detectors are
provided easily and with low cost.
The thin NaI(Tl) plate has been developed for this project by 
Horiba Ltd.\@ successfully with low cost and with high purity.
The NaI(Tl) array needs the large number of channels for data 
acquisition (DAQ) system.
Recently, DAQ systems for the large number of module (1024-2048 channels)
are extensively developed in the fields of high energy experiment.
The low noise measurement should be performed for the present work.
Since the developed thin NaI(Tl) has the low energy threshold ($\sim$2keV),
the noise from PMT will not be a big problem.
However, the pulse shape analysis will be performed in order to monitor the 
PMT noises.

We showed that the thin NaI(Tl) array has excellent sensitivity for 
WIMP-nucleus inelastic excitation by only 10kg of NaI(Tl).
The highly segmented detector system has the great advantage to 
SD-type WIMPs candidates.

\section{Acknowledgment}
The authors thank Prof. H.Ejiri, Prof. M.Nomachi and Prof. T.Kishimoto 
for fruitful discussions.
They also thank Prof. A.Masaike for his encouragement of our work.
The present work was supported by Toray Science Foundation.

\end{document}